\documentclass{cimento}

\usepackage{graphicx}  

\newcommand\POWHEG{{\tt POWHEG}}
\newcommand\MCatNLO{{\tt MC@NLO}}
\newcommand\MiNLO{{\tt MiNLO}}
\def\sss{\scriptscriptstyle}
\newcommand\as{\alpha_{\sss\rm S}}
\newcommand\Phirad{\Phi_{\mathrm{rad}}}
\def\nn{\nonumber}
\usepackage{amssymb}
\usepackage{amsmath}
\newcommand{\noun}[1]{\textsc{#1}}
\newcommand{\HNNLO}{\noun{Hnnlo}}
\newcommand{\HQT}{\noun{HqT}}
\newcommand{\JETVHETO}{\noun{JetVHeto}}

\title{Reaching NNLOPS accuracy with POWHEG and MiNLO}
\author{E.~Re}
\instlist{\inst{} Rudolf Peierls Centre for Theoretical Physics, Department
  of Physics,\\University of Oxford, Oxford, OX1 3NP, United Kingdom}

\PACSes{\PACSit{12.38.Bx}{Perturbative calculations} \PACSit{14.80.Bn}{Standard-model Higgs bosons}}

\begin{document}

\maketitle

\begin{abstract}
  We describe how a simulation of Higgs boson production accurate at
  next-to-next-to-leading order and matched to a parton shower can be
  built by combining the \POWHEG{} and \MiNLO{} methods and using
  \HNNLO{} results as input.
\end{abstract}

\section{Introduction}

During the last decade a major research effort in the Monte Carlo
community has been devoted to the development of NLOPS tools,
\emph{i.e.} tools that allow a matching of next-to-leading order (NLO)
computations with parton showers (PS), thereby bringing NLO accuracy
into standard Monte Carlo event
generators~\cite{Buckley:2011ms}. Among many proposals, there are
currently two well-established NLOPS approaches, namely
\POWHEG{}~\cite{Nason:2004rx,Frixione:2007vw} and
\MCatNLO{}~\cite{Frixione:2002ik}, which have now become the methods
of choice used by experimental collaborations in many searches being
carried out at the LHC. Part of this success was possible due to the
progress in the automation of NLO computations,
in the development of semiautomated or fully-automated NLOPS
frameworks~\cite{Alioli:2010xd,Frederix:2011zi,Platzer:2011bc,Hoeche:2011fd},
as well as in the standardization of well-defined
interfaces~\cite{Binoth:2010xt,Alioli:2013nda} between programs that
operate different tasks.

A topic of research that has received much attention during the last 2
years is the merging of multiple NLOPS simulations for different jet
multiplicities. These advances represent the NLO generalization of
well-established tree-level multileg merging
approaches~\cite{Catani:2001cc,Mangano:2006rw}, and their relevance
for future LHC phenomenology is clear, since they will allow a
significant improvement in the simulation of processes where a heavy
system is produced in association with multiple jets, which is the
generic background for many new-Physics searches. There have been
several proposals aiming at this
goal~\cite{Alioli:2011nr,Hoeche:2012yf,Frederix:2012ps,Platzer:2012bs,
  Alioli:2012fc, Lonnblad:2012ix,Hamilton:2012rf,Hartgring:2013jma},
among which the \MiNLO{}
approach~\cite{Hamilton:2012np,Hamilton:2012rf}.

After a short review of the \POWHEG{} and \MiNLO{} approaches, I will
describe how their combination can be used to match NNLO computations
with PS, and show recent results obtained for inclusive Higgs
production~\cite{Hamilton:2013fea}.

\subsection{\POWHEG{}}
The \POWHEG{} method is a prescription to interface NLO calculations
with parton shower generators avoiding double counting of real
emissions and virtual corrections.
In the \POWHEG{} formalism, the generation of the hardest emission is
performed first, according to the distribution given by
\begin{equation}
\label{eq:master}
d\sigma=\bar{B}\left(\Phi_{B}\right)\,d\Phi_{B}\,\left[\Delta_{R}\left(p_{T}^{\min}\right)+
\frac{R\left(\Phi_{R}\right)}{B\left(\Phi_{B}\right)}\,\Delta_{R}\left(k_{T}\left(\Phi_{R}\right)\right)\,d\Phi_{\mathrm{rad}}\right]\,,
\end{equation}
where $B\left(\Phi_{B}\right)$ is the leading order contribution,
\begin{equation}
\label{eq:bbar}
\bar{B}\left(\Phi_{B}\right)=B\left(\Phi_{B}\right)+
\left[V\left(\Phi_{B}\right)+\int d\Phi_{\mathrm{rad}}\, R\left(\Phi_{R}\right)\right]
\end{equation}
is the NLO differential cross section integrated on the radiation
variables while keeping the Born kinematics fixed
($V\left(\Phi_{B}\right)$ and $R\left(\Phi_{R}\right)$ stand
respectively for the virtual and the real corrections), and
$\Delta_{R}\left(p_{T}\right)=\exp\left[-\int d\Phi_{\mathrm{rad}}\,\frac{R\left(\Phi_{R}\right)}{B\left(\Phi_{B}\right)}\,\theta\left(k_{T}\left(\Phi_{R}\right)-p_{T}\right)\right]\,$
is the \POWHEG{} Sudakov. With $k_T\left(\Phi_{R}\right)$ we denote
the transverse momentum of the emitted particle off a Born-like
kinematics $\Phi_B$, and, as usual, the cancellation of soft and
collinear singularities is understood in the expression within the
square bracket in eq.~(\ref{eq:bbar}).  Partonic events with hardest
emission generated according to eq.~(\ref{eq:master}) are then
showered with a $k_T$-veto on following emissions.  Subject to these
conditions, it can be shown that such events
exhibit the features typical of PS when the chosen observable probes
the soft-collinear regions (Sudakov suppression), reproduce the exact
fixed-order results in the regions where emissions are widely
separated, and, crucially, they preserve NLO accuracy for inclusive
observables. From the NLOPS-matching point of view,
the more challenging processes currently described with this approach
are $2\to 3$ and $2\to 4$ processes, with at most 2 light jets at
LO~\cite{Campbell:2012am,Re:2012zi,Jager:2012xk,Jager:2013mu}.  

For the benefit of the following discussion, the (unregulated)
$\bar{B}$ function of the standard \POWHEG{} simulation of $H+1$ jet
can be written schematically as
\begin{equation}
  \label{eq:bbarH1j}
  \bar{B}_{\,\tt HJ}=\as^3(\mu_R) \Big{[} B + \as V(\mu_R) + \as \int d\Phirad R \Big{]}\,,
\end{equation}
where we have made explicit the dependence of all terms upon $\as$ and
the renormalization scale $\mu_R$. It is also worth recalling that
when one or more jets are present at LO (as in the $H+1$ jet case) the
associated $\bar{B}$ function needs to be regulated from the
divergences arising when jets in the LO kinematics become
unresolved~\cite{Alioli:2010qp}: as a consequence, a standard
\POWHEG{} simulation of $H+1$ jet cannot be used to describe inclusive
Higgs production.

\subsection{\MiNLO{}}
It is known that a common issue present in multileg NLO computations
is the choice of the factorization ($\mu_F$) and renormalization
scale: ultimately the problem is due to the fact that these
computations
are characterized by kinematical regimes involving 
several different scales, and, although some choices are clearly pathologic
(as they can lead for instance to negative cross sections), in general
there is no procedure to a-priori choose $\mu_R$ and $\mu_F$, being
the scale dependence of the result just an artefact of truncating the
perturbative expansion.

The \MiNLO{} procedure~\cite{Hamilton:2012np} was originally defined
as a prescription to address this issue, and it works by consistently
including CKKW-like corrections into a standard NLO computation.
By clustering with a $k_T$-measure the momenta of each phase-space
point occurring in the computation, one can define the
``most-probable'' branching history that would have produced such a
kinematics: the argument of each power of $\as$ is then found from the
transverse momentum of the splitting occurring at each nodal point of
the skeleton built from clustering, and a prescription for $\mu_F$ is
given as well. The result is also corrected by means of Sudakov form
factors (called \MiNLO{}-Sudakov FF's in the following) associated to
internal lines, accounting for the large logarithms that arise when
the clustered event contains well separated scales.

Because of the presence of \MiNLO{}-Sudakov FF's associated to the
Born-like kinematics, the integration over the full phase space
$\Phi_B$ can be performed without generation cuts: a \MiNLO{}-improved
computation yields finite results also when jets in the LO kinematics
become unresolved. As a consequence, the \MiNLO{} procedure can be
used within the \POWHEG{} formalism to regulate the $\bar{B}$ function
for processes involving jets at LO, without using external cuts or
variants thereof.

\MiNLO{}-enhanced \POWHEG{} simulations have been presented in
refs.~\cite{Hamilton:2012np,Hamilton:2012rf,Campbell:2013vha,Luisoni:2013cuh}
and, in particular, in the $H+1$ jet case, the master formula for
generating the hardest emission contains the following $\bar{B}$
function
\begin{eqnarray}
  \label{eq:bbarH1jMiNLO}
  \bar{B}_{\,\tt HJ-MiNLO}&=&\as^2(M_H) \as(q_T) \Delta^2_g(q_T,M_H)\\
  &\times&\Big{[}
  B ( 1-2\Delta^{(1)}_g(q_T,M_H) ) + 
  \as V(\bar{\mu}_R) +\as \int d\Phirad R
  \Big{]}\nn\,,
\end{eqnarray}
that should be contrasted with eq.~(\ref{eq:bbarH1j}). In
eq.~(\ref{eq:bbarH1jMiNLO}) $q_T$ is the Higgs transverse momentum (in
the underlying-Born kinematics), $M_H$ is its virtuality,
$\bar{\mu}_R$ is set to $(M_H^2 q_T)^{1/3}$ in accordance with the
\MiNLO{} prescription and
$\Delta_g(q_T,Q)=\exp\Big{\{}-\int_{q^2_T}^{Q^2}\frac{dq^2}{q^2}\frac{\as(q^2)}{2\pi}
\Big{[} A_{g}\log\frac{Q^2}{q^2} + B_{g} \Big{]}\Big{\}}$ is the
\MiNLO{}-Sudakov FF associated to the jet present at LO. At NLL, the
$A_{1,g}$, $A_{2,g}$ and $B_{1,g}$ terms in the expansion of $A_g$ and
$B_g$ need to be included~\cite{Hamilton:2012np}. The term in
brackets multiplying $B$ is needed to avoid double-counting of NLO
factors: $ \Delta_{g}^{(1)}(q_T,Q) = - \frac{\as}{2\pi} \Big{[}
\frac{1}{2}A_{1,g}\log^2\frac{Q^2}{q_T^2} + B_{1,g}
\log\frac{Q^2}{q_T^2} \Big{]}$ corresponds to the $\mathcal{O} (\as)$
expansion of $\Delta_g$.

The $\bar{B}$ function in eq.~(\ref{eq:bbarH1jMiNLO}) can be
integrated over the full phase space associated with the ``LO'' jet,
yielding a finite cross-section for inclusive Higgs production. The
formal accuracy of the result so obtained was carefully addressed
in ref.~\cite{Hamilton:2012rf}, by means of a comparison with the NNLL
$q_T$-resummation of the Higgs transverse momentum. It was found that,
in order to reach NLO accuracy for the total inclusive Higgs
production, the NNLL $B_{2,g}$ term should be included in the
\MiNLO{}-Sudakov FF, and $q_T$ should be used as factorization scale
and as the argument of the power of $\as$ associated to $R$, $V$ and
$\Delta_{g}^{(1)}$ (\emph{i.e.} the power of $\as$ where no argument
was specified in eq.~(\ref{eq:bbarH1jMiNLO})).  If such terms are not
included properly, spurious terms of order $\as^{3.5}$ are generated
upon integration over the entire Higgs $p_T$ spectrum, violating the
requirement $[d\sigma_{\tt HJ-MiNLO}]_{\rm integrated}-\sigma_{\rm NLO
}(gg\to H)=\mathcal{O}(\as^4)$, which is needed to claim NLO accuracy
for fully-inclusive Higgs production.

\section{Higgs production with NNLOPS accuracy}
The $H+1$ jet \POWHEG{} implementation enhanced with the improved
\MiNLO{} procedure previously outlined can be used to reach NNLOPS
accuracy. In fact, since such a simulation gives a NLO-accurate
prediction of the Higgs rapidity ($y$), then the function $W(y)$,
defined as
\begin{equation}
  \label{eq:wy}
  W(y)=\frac{(d\sigma/dy)_{\rm NNLO}}
  {(d\sigma/dy)_{\tt HJ-MiNLO}}\,,
\end{equation}
can be used to reweight each {\tt HJ-MiNLO}-generated event, thereby
obtaining a NNLOPS simulation of inclusive Higgs production.  By
NNLOPS we mean a fully-exclusive Monte Carlo simulation of
Higgs-production which is NNLO accurate when one is fully inclusive on
extra radiation, as well as LO (NLO) accurate for $H+2 (1)$ jet
observables~\cite{Hamilton:2012rf,Hamilton:2013fea}. Since we are
reweighting with $W$, the Higgs rapidity is NNLO accurate by construction,
whereas the NLO accuracy of the 1-jet region, inherited from the
underlying {\tt HJ-MiNLO} simulation, is not spoiled, because the
first non-controlled terms in the whole simulation are
$\mathcal{O}(\as^5)$: this follows from the fact that $W(y)=1+
\mathcal{O}(\as^2)$, as can be seen expanding numerator and
denominator in eq.~(\ref{eq:wy}).

In ref.~\cite{Hamilton:2013fea} the following generalization of
eq.~(\ref{eq:wy}) was used:
\begin{equation}
  \label{eq:wypaper}
  W(y,p_T)=h(p_T)\frac{\int d\sigma_{\rm NNLO}\delta(y-y(\Phi)) - \int d\sigma_{\tt HJ-MiNLO}^B \delta(y-y( \Phi)) }{\int d\sigma_{\tt HJ-MiNLO}^A \delta(y-y(\Phi)) } + (1-h(p_T))\,,
\end{equation}
where we have split the {\tt HJ-MiNLO} differential cross section
among $d\sigma^A = d\sigma\ h(p_T)$ and $d\sigma^B=d\sigma\
(1-h(p_T))$, with $h(p_T)=\frac{(\beta m_H)^2}{(\beta
  m_H)^2+p_T^2}$. The profiling function $h$ controls where the
NLO-to-NNLO correction is spread: as $2^{\mbox{\scriptsize{nd}}}$
argument of $W$ the transverse momentum of the leading jet was used,
and we have chosen $\beta=1/2$, which implies that the NNLO correcting
factor $W$ is effectively applied in the region $p_T\lesssim m_H/2$
(for $p_T\gg m_H$, $W(y,p_T)\to 1$).  With the choice in
eq.~(\ref{eq:wypaper}) one also has that $(d\sigma/dy)_{\rm NNLOPS}$
reproduces $(d\sigma/dy)_{\rm NNLO}$ exactly, without $\mathcal{O}
(\as^5)$ ambiguities.

\subsection{Results}
In our simulation, the central value for $d\sigma_{\rm NNLO}$ was
obtained with \HNNLO{}~\cite{Catani:2007vq,Grazzini:2008tf}, setting
$\mu_R=\mu_F=m_H/2$. We refer to ref.~\cite{Hamilton:2013fea} for
details on how scales were varied to obtain uncertainty bands.

\begin{figure}[!htb]
  \includegraphics[scale=0.6]{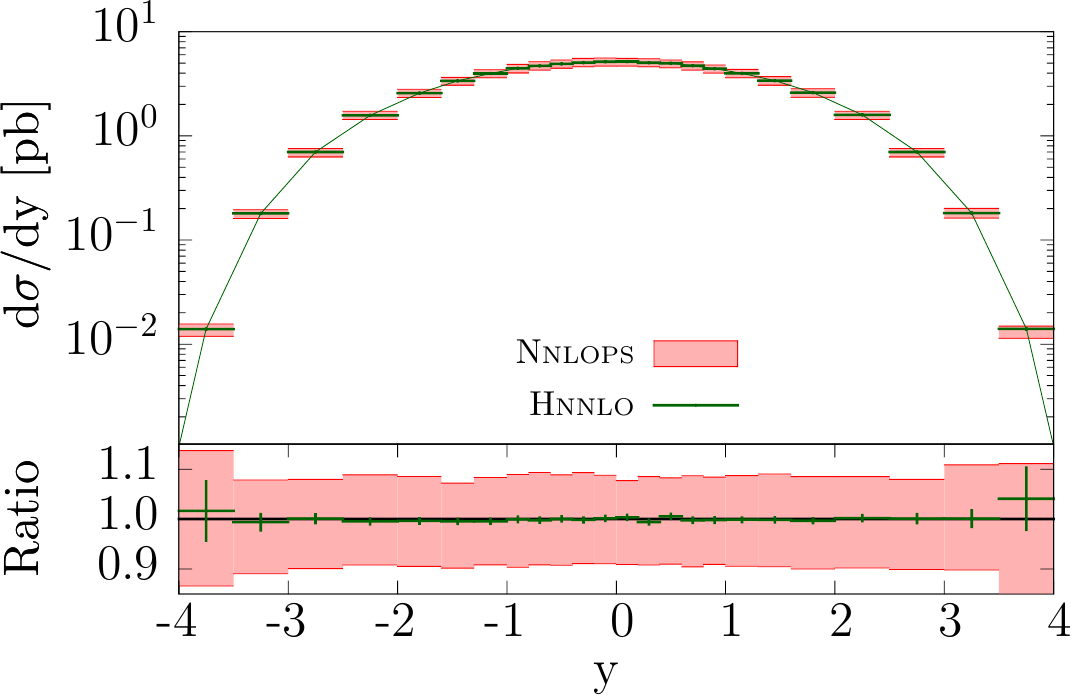}~
  \includegraphics[scale=0.6]{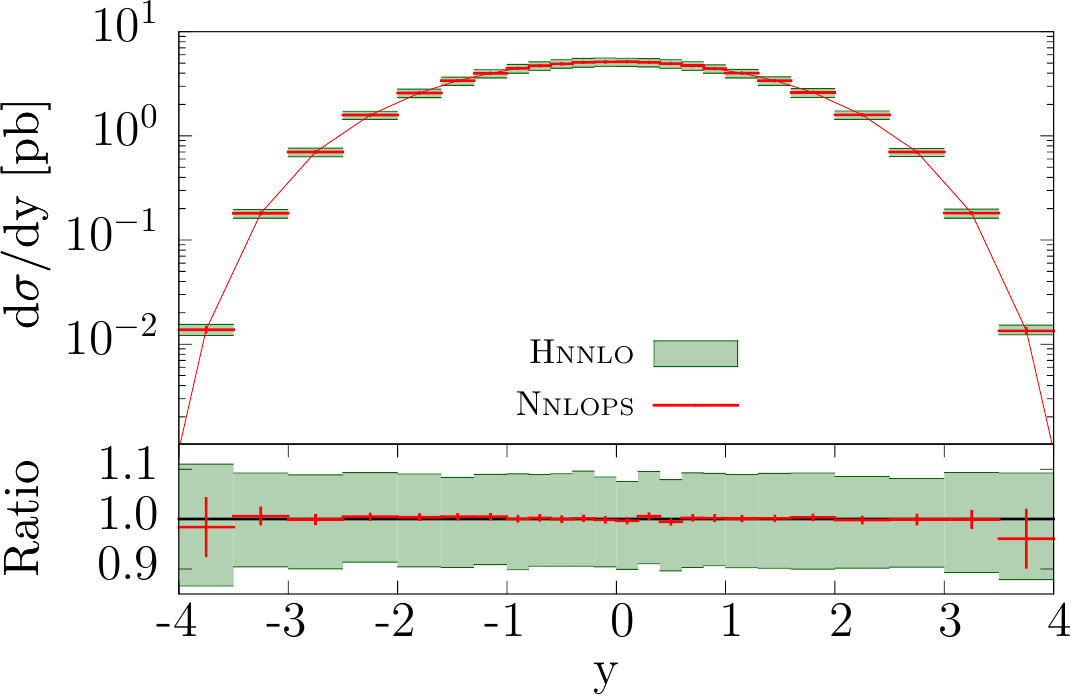}
  \caption{Comparison of the NNLOPS (red) and \HNNLO{} (green) results
    for the Higgs fully inclusive rapidity distribution. On the left
    (right) plot only the NNLOPS (\HNNLO{}) uncertainty is displayed.
    The lower left (right) panel shows the ratio with respect to the
    NNLOPS (\HNNLO{}) prediction obtained with its central scale
    choice.}
  \label{fig:y}
\end{figure}

In fig.~\ref{fig:y} a comparison between our NNLOPS simulation and
\HNNLO{} is shown: as expected, the NNLOPS simulation reproduces
extremely well the NNLO results for the Higgs rapidity both in the
central value and in the uncertainty band obtained by scale variation.

\begin{figure}[!htb]
  \includegraphics[scale=0.6]{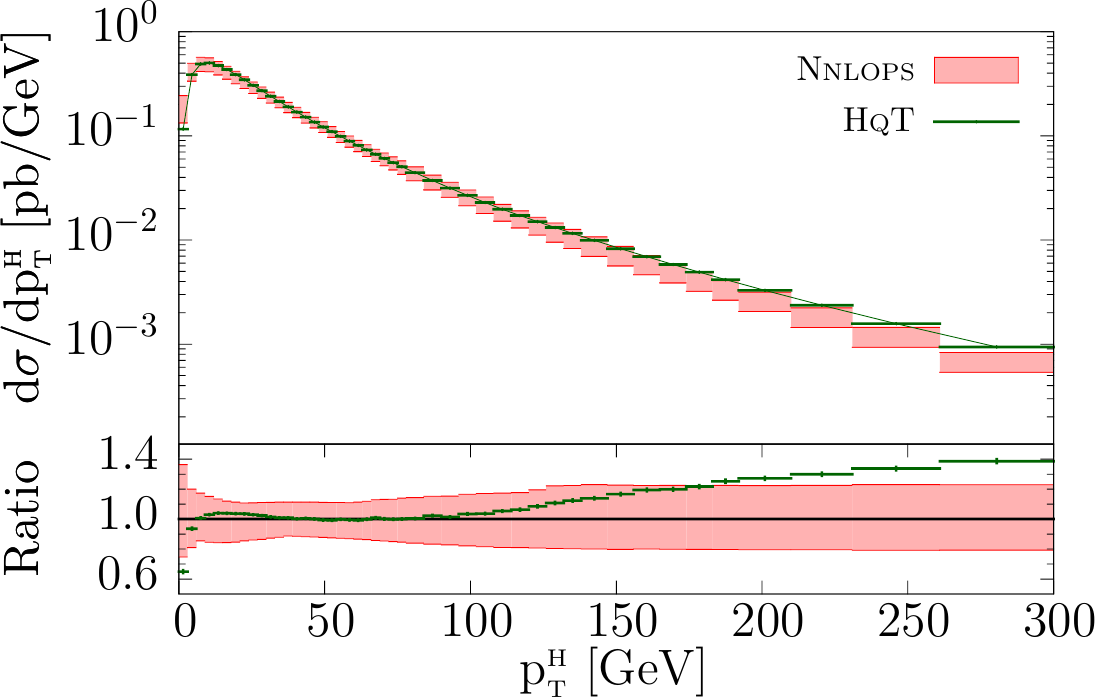}~
  \includegraphics[scale=0.6]{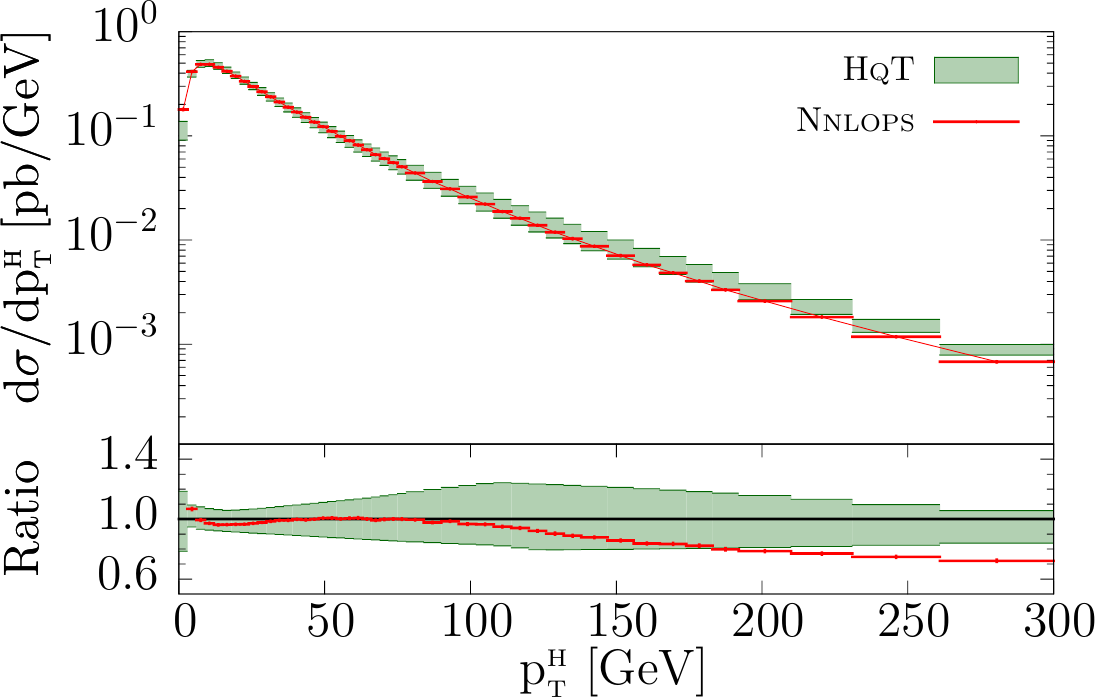}
  \caption{Comparison of the NNLOPS (red) with the NNLL+NNLO
    prediction of \HQT{} (green) for the Higgs transverse momentum.
    In \HQT{} we keep the resummation scale $Q_{\rm res}$ always
    fixed to $m_H/2$ and vary $\mu_R$ and $\mu_F$.  On the left
    (right), the NNLOPS (\HQT{}) uncertainty band is shown. In the
    lower panel, the ratio to the NNLOPS (\HQT{}) central
    prediction is displayed.}
  \label{fig:pth}
\end{figure}

Fig.~\ref{fig:pth} shows the Higgs transverse momentum $p_T^H$. We
compare our simulation with
\HQT{}~\cite{Bozzi:2005wk,deFlorian:2011xf}, whose central value is
obtained with $Q_{\rm res}=m_H/2$ and $\mu_R=\mu_F=m_H/2$. The \HQT{}
result corresponds to a NNLL prediction of $p_T^H$, matched to the
fully inclusive cross section at NNLO. Here we notice that the two
results are almost completely contained within each other's
uncertainty band in the region of low-to-moderate transverse
momenta. The central values at small momenta also exhibit a very good
agreement, supporting our choice for $\beta$. The difference in the
large-$p_T$ tail is not a reason of concern, and it is expected since
the two predictions use different scales at large $p_T$, as explained
in ref.~\cite{Hamilton:2013fea}.

Finally, we also mention that a comparison among NNLOPS and NNLL+NNLO
predictions from \JETVHETO{}~\cite{Banfi:2012jm} was successfully
carried out for the jet veto efficiency, defined as the cross section
for Higgs boson production events containing no jets with transverse
momentum greater than a given value ($p_{\scriptscriptstyle
  \mathrm{T,veto}}$), divided by the respective total inclusive cross
section. The central predictions of the two programs are never out of
agreement by more than 5-6\%, and the two sets of predictions lie
within each other's error bands essentially everywhere over all values
of $p_{\scriptscriptstyle \mathrm{T,veto}}$, as shown in
ref.~\cite{Hamilton:2013fea}.

\acknowledgments NNLOPS results presented here have been obtained in
ref.~\cite{Hamilton:2013fea}, in collaboration with K.~Hamilton,
P.~Nason and G.~Zanderighi. The original proposal of reaching NNLOPS
accuracy from \MiNLO{}-merged NLOPS simulations was outlined in
ref.~\cite{Hamilton:2012rf}, which was co-authored by C.~Oleari. The
author acknowledges G.~Corcella and L.~Pancheri for the invitation to
the LC13 workshop in Trento, and the ``HadronPhysics3'' project for
covering part of the associated living expenses.

\end{document}